\documentclass[pre,twocolumn,aps,showpacs]{revtex4}
\usepackage{dcolumn}
\usepackage{graphicx}
\usepackage{graphicx,amssymb,amsmath}
\usepackage{natbib}
\usepackage{bm}

\newcommand{\affmeas}{m}

\begin{document}

\title{Distinct regimes of elastic response and deformation modes
of cross-linked cytoskeletal and semiflexible polymer networks}

\author{D.A. Head$^{1,2}$, A.J. Levine$^{2,3}$
and F.C. MacKintosh$^{1,2}$}

\affiliation{$^{1}$Department of Physics and Astronomy,
Vrije Universiteit, Amsterdam, The Netherlands.\\
$^{2}$Kavli Institute for Theoretical Physics,
University of California at Santa Barbara CA 93106, USA.\\
$^{2}$Department of Physics, University of Massachusetts,
Amherst MA 01060, USA.}

\date{\today}

\begin{abstract}
Semiflexible polymers such as filamentous actin play a
vital role in the mechanical behavior of cells, yet the basic
properties of cross--linked F--actin networks remain poorly
understood. To address this issue, we have performed numerical 
studies of the linear response of
homogeneous and isotropic two--dimensional networks subject to an applied strain at zero temperature.
The elastic moduli are found to vanish for network densities at a rigidity percolation threshold.
For higher densities, two regimes are observed: one in which the deformation is predominately affine
and the filaments stretch and compress; and a second in which bending modes dominate.
We identify a dimensionless scalar quantity,
being a combination of the material length scales,
that specifies to which regime a given network
belongs. A scaling argument is presented that approximately
agrees with this crossover variable.
By a direct geometric measure, we also confirm that the 
degree of affinity under strain correlates with the 
distinct elastic regimes. 
We discuss the implications of our findings and suggest
possible directions for future investigations.
\end{abstract}

\pacs{87.16.Ka, 62.20.Dc, 82.35.Pq}

\maketitle

\section{Introduction}

The mechanical stability, response to stress, and
locomotion of eukaryotic cells is largely due to networks
of biopolymers that collectively form what is known as
the cytoskeleton
\cite{alberts:88,elson:88,janmey:90,actin_gels,sackmann:02}. Filamentous actin (F--actin), microtubules and other intermediate filamentous proteins make up the cytoskeletal
network, along with a variety of auxiliary proteins that govern such factors as cross-linking and filament growth.
By understanding the relation of the individual filament properties and dynamically evolving gel microstructure to the rheological/mechanical properties of intracellular structures, one will better understand the general framework for cellular force generation and transduction~\cite{elson:88,wang:93,bohm:97,stamenovic:00}. Such stress production and sensing underlies such fundamental biological processes as cell division, motility~\cite{stossel:93}, and adhesion~\cite{mazia:75,braun:98,guttenberg:01,ralphs:02}. Given the importance of biopolymer networks in determining the mechanical response of cells, there is an obvious interest in understanding the properties of such networks at a basic level. Understanding stress propagation in cells also has implications for the interpretation of intracellular microrheology experiments~\cite{yamada:00,fabry:01}. However, biopolymers also belong to the class of
semiflexible polymers, so--called because their characteristic bending length (however defined) is
comparable to other length scales in the problem, such as the contour length or the network mesh size,
and thus cannot be neglected.

Such semiflexible polymers pose interesting and fundamental challenges in their own right as polymer materials. The understanding of the properties of individual semiflexible polymers is quite highly developed~\cite{fixman:73,schmidt:89,smith:92,farge:93,gittes:93,ott:93,wilhelm:96,granek:97,gittes:98,caspi:98,everaers:99}; in addition, the dynamical and rheological 
properties of these polymers in solution~ \cite{odijk:83,semenov:86,ruddies:93,isambert:96,morse:98,gittes:98,hinner:98,morse:99-01} have largely been elucidated. The remaining problem of determining the rheology of permanently cross-linked gels of semiflexible polymers, however, has proved quite subtle. 
Related theoretical approaches considered thus far
have either assumed a simplifying network
geometry, such as a lattice~\cite{satcher96}
or a Cayley tree~\cite{jones_ball},
or assumed that the dominant deformation modes
are affine~\cite{mackintosh:95}
or dominated by transverse filament fluctuations 
and bending~\cite{kroy:96,frey:98}.

In this paper we study the static mechanical properties of random, semiflexible, cross-linked networks in the linear response regime with the aim of shedding light on the more complex, nonequilibrium cytoskeleton. Our approach is deliberately minimalistic: we consider two--dimensional, athermal systems with no polydispersity
in filament properties. Although obviously simplified, this restricts the parameter
space to a manageable size and allows for a fuller characterization of the network
response. The central finding of our work is the existence of qualitatively distinct regimes in the elastic response and  local deformation in networks, each with characteristic signatures that should be observable experimentally.

The basic distinction between stress propagation in flexible and semiflexible networks is that in the former elastic deformation energy is stored entropically in the reduction of the number of chain conformations between cross-links~\cite{deGennes:79,rubenstein:03} while in the latter, the elastic energy is stored primarily in the mechanical bending and stretching of individual chains. In a flexible, cross-linked mesh there is only one microscopic length scale: the mean distance between cross-links or entanglements, $l_{\rm c}$. Since the actual identity of a flexible chain is immaterial on scales beyond this mean cross-linking distance, chain length plays only a very small role~\cite{ends}. In a semiflexible gel, however, chains retain their identity through a cross-link because the tangent vectors on a given chain remain correlated over distances much longer than $\xi$, the mesh size. Thus, it is reasonable to expect that the elastic properties of the network depend on both the mesh size and the length of the chains.

\begin{figure}[ht]\centering
\includegraphics[width=8cm]{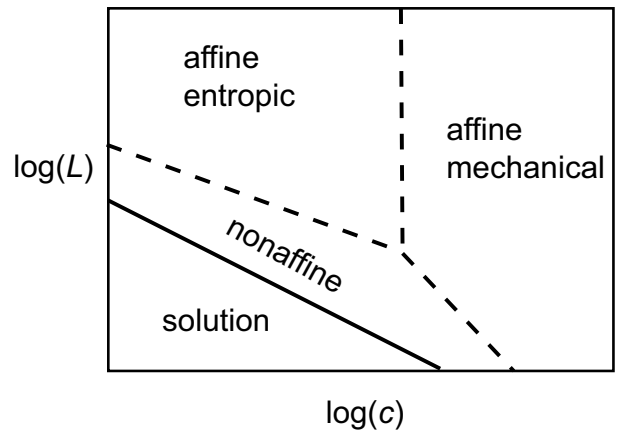}
\caption{A sketch of the expected diagram showing the various
elastic regimes in terms of molecular weight $L$ and concentration
$c\sim1/l_{\rm c}$. The solid line represents the rigidity percolation
transition where rigidity first develops at a macroscopic level. 
This transition is given by $L\sim c^{-1}$. The other, dashed lines
indicate crossovers (not thermodynamic transitions), as described in the text. 
As sketched here, the crossovers between nonaffine and affine 
regimes demonstrate the independent nature of these crossovers from the rigidity percolation transition.}
\label{f:PD}
\end{figure}

Upon increasing the density of filaments and thus the density of cross-links (hereafter we refer to filament density only as one quantity determines the other in 2d) the system acquires a static shear modulus via a continuous phase transition at the rigidity percolation point. This corresponds to moving from left to right in the lower half of the phase diagram shown in Fig.\ \ref{f:PD}. This critical point (solid line in Fig.\ \ref{f:PD}) is at higher density than the connectivity percolation point. Since our model ignores the entropic elasticity of the network, below this critical density associated with rigidity percolation, the material has no static shear modulus and may be considered a liquid. We will discuss the finite temperature implications of this zero--temperature phase transition in more detail. The elastic properties of the fragile gel (solid) that exists just above this critical point are controlled by the physics of rigidity percolation. 

By increasing the cross-link density further we encounter a regime over which the elastic deformation of the gel is dominated by filament bending and is highly nonaffine. This \emph{nonaffine} regime (NA) is consistent with prior predictions by Kroy and Frey~\cite{kroy:96,frey:98}. Within the NA regime the static shear modulus scales linearly with the bending modulus of the individual filaments, $\kappa$, but the elastic moduli are not controlled by properties of the rigidity percolation critical point. Most remarkably, however, the deformation field under uniformly applied stress is highly heterogeneous spatially over long length scales comparable even to the system size. We will quantify the degree of nonaffinity as a function of length scale and use this nonaffinity measure to demonstrate that the degree of nonaffinity in the NA regime increases without bound as one goes to progressively smaller length scales. Such materials are then poorly described by standard continuum elasticity theory on small length scales.

By further increasing the filament density, one approaches a crossover to a regime of affine deformation (A), in which the strain is uniform throughout the sample, as the velocity field would be in a simple liquid under shear. This cross-over is shown in Fig.\ \ref{f:PD} by the dashed lines above the NA region. In this regime, elastic energy is primarily stored in the extension/contraction of filaments. Also, in contrast to the NA regime, the growth of the degree of nonaffinity saturates as a function of decreasing length scale. The 
elastic response of the network is governed primarily by the longitudinal compliance of filaments, and the shear 
modulus can be calculated from the combination of this realization and the assumption of affine, uniform deformation as shown in Refs.~\cite{mackintosh:95,isambert:96,morse:98,hinner:98,gittes:98,morse:99-01}.  We will show that there is one dimensionless parameter $\lambda$ that controls the $NA \rightarrow A$ crossover. It is set by the ratio of the filament length to a combination of parameters describing the density of the network and the individual filament stiffness. 

In the affine regime, the longitudinal response actually can arise from two distinct mechanisms: there are two forms of compliance of a semiflexible filament under extensional stress, one essentially entropic \cite{mackintosh:95} and the other essentially mechanical \cite{frey:98}. In the first case, the compliance relates to the thermally fluctuating filament conformation, which, for instance, is straightened-out under tension. A change in the length of a filament between cross-links results not in simple mechanical strain along that filament but rather in a reduction of the population of transverse thermal fluctuations along that filament thereby reducing the entropy of the filament.  This reduction results in an elastic restoring force along the length of the filament. This is the dominant compliance for long enough filament segments (\emph{e.g.}, between cross--links). In the second case, the compliance is due to a change in the contour length of the filament under tension, which, although small, may dominate for short segments (\emph{e.g.}, at high concentration). Thus, in general, we find two distinct affine (A) regimes, which we refer to as entropic (AE) and mechanical (AM). 

Moving still further up and to the left in the diagram shown in Fig.~\ref{f:PD} we would eventually reach a regime (not shown) in which the filament lengths between cross-links are much longer than their thermal persistence length and standard rubber elasticity theory would apply. This regime is of no experimental importance for the actin system.  To complete our description of the diagram, we note that by increasing the filament concentration in the entropic affine regime, one must find a transition from entropic to mechanical elasticity ($AE \longrightarrow AM$) within a regime of affine deformation.

Here we confine our attention to two-dimensional permanently cross-linked networks and consider only enthalpic contributions to the elastic moduli of the system. In effect we are considering a zero-temperature system, except that we account for the extensional modulus of F-actin that is principally due to the change in the thermal population of transverse thermal fluctuations of a filament under extension. We do not expect there to be a significant entropic contribution to the free energy coming from longer length scale filament contour fluctuations because of their inherent stiffness. This justifies our neglect of the sort of entropic contributions to the filament free energy that are typically considered in the analysis of rubber elasticity of flexible polymers. We discuss this more below. The low dimensionality of the model system, on the other hand, is more significant since the essentially straight chains in two dimensions will not interact sterically under small deformations. In our model, only one length scale is required to describe the random network, the mean distance between cross-links, $l_{\rm c}$. In three dimensions, however, chains can interact sterically and two quantities are needed to fully describe the random network, the density of cross-links and density of filaments. 

In our zero-temperature analysis of the system presented in this paper, we do not explicitly probe the difference between the $NA\longrightarrow AE$ and the $NA\longrightarrow AM$ cross-overs. Their difference enters our description of the system via a choice made for the form of $\mu$, the extension modulus of an individual filament. For physiological actin, we expect that the relevant transition will be the $NA\longrightarrow AE$. We discuss this further in section \ref{s:thermal}.

The remainder of this paper is arranged as follows. In Sec.\ \ref{s:model} we define our model system
in terms of the mechanical properties of individual filaments and the manner in which they interconnect
to form the network. An overview of the simulation method used to find the mechanical equilibrium under an imposed strain is also given. We describe in Sec.\ \ref{s:percolation} the rigidity percolation transition, at which network rigidity first develops. We then describe the crossover from a nonaffine regime
above the rigidity transition to an affine regime in Sec.\ \ref{s:elastic}. A scaling argument is also presented for 
this crossover. The macroscopic mechanical response is also demonstrated to be linked to geometric measures of the degree of affine deformation at a local level. In Sec.~\ref{s:thermal}, we show how the network response in the affine regime can be either essentially thermal or mechanical in nature. In Sec.~\ref{s:discuss}, we discuss primarily the experimental implications of our results.

\section{The model}
\label{s:model}

The bending of semiflexible polymers has been successfully described by the wormlike chain model, in which non--zero
curvatures induce an energy cost according to a bending modulus $\kappa$. For small curvatures, the Hamiltonian can be written
\begin{equation}
{\mathcal H}_{\rm bend}
=
\frac{1}{2}\kappa
\int{}{\rm ds}(\nabla^{2}{\bf u})^{2}
\label{e:H_bend}
\end{equation}
\noindent{}where ${\bf u}(s)$ is the transverse displacement of the filament, and $s$ is integrated along the total contour length of the filament. Transverse filament dynamics can be inferred from this
Hamiltonian~\cite{farge:93}. For finite temperatures, (\ref{e:H_bend}) can also be used
to predict the longitudinal response of an isolated filament; however, as $T\rightarrow0$ the filament buckles,
preconfiguring a breakdown of the linear response~\cite{frey:98}. We also consider the response of a filament to compression/extensional deformations through the elastic Hamiltonian given by:
\begin{equation}
{\mathcal H}_{\rm stretch}
=
\frac{1}{2}
\mu
\int{\rm d}s
\left(
\frac{{\rm d}l(s)}{{\rm d}s}
\right)^{2}
\label{e:H_stretch}
\end{equation}
\noindent{}where ${\rm d}l/{\rm d}s$ gives the relative change in length along the filament.
(\ref{e:H_stretch}) is just a Hookean spring response, with a stretching modulus $\mu$ that
is here taken to be independent of $\kappa$, although they can both be related to the cross--sectional radius and elastic properties of individual filaments (see Sec.~\ref{s:scaling}).

\begin{figure}[ht] \centering
\includegraphics[width=8cm]{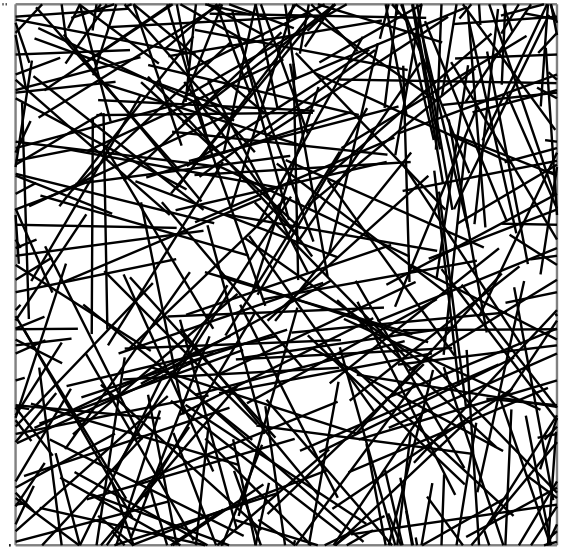}
\caption{An example of a network with a cross--link
density $L/l_{\rm c}\approx29.09$       
in a shear cell of dimensions $W\times W$ and
periodic boundary conditions in both directions.
This example is small, $W=\frac{5}{2}L$;
more typical sizes are $W=5L$ to $20L$.
}
\label{f:eg_geo}
\end{figure}

The networks are constructed by the sequential random deposition of monodisperse filaments of length $L$ into a two--dimensional shear cell with dimensions $W\times W$. Since the position and orientation of filaments are uniformly distributed over the allowed ranges, the networks are isotropic and homogeneous when viewed on sufficiently large length scales. Each intersection between filaments is identified as a cross--link, the mean distance between which (as measured along a filament) is denoted $l_{\rm c}$\,,
so that the mean number of cross--links per rod
is $L/l_{\rm c}-1$. Deposition continues until the required cross-linking density $L/l_{\rm c}$ has been reached.
An example network geometry is given in Fig.~\ref{f:eg_geo}.

\begin{figure}
\includegraphics[width=8cm]{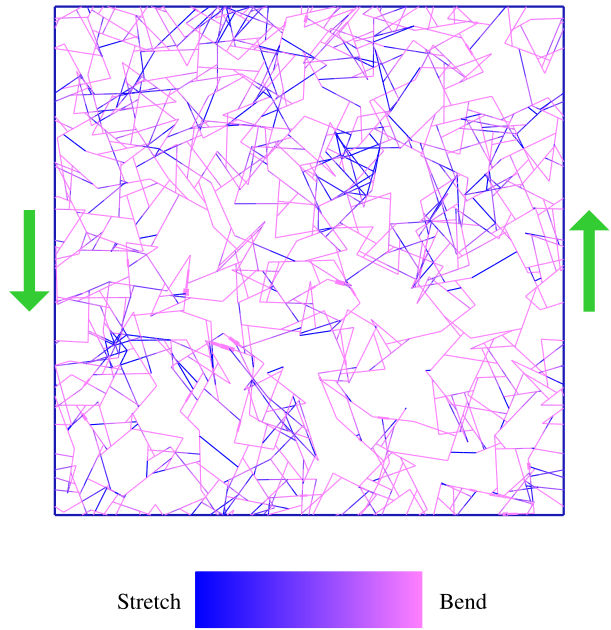}
\caption{{\em Color online}
An examples of a low--density
network with $L/l_{\rm c}\approx8.99$ in mechanical equilibrium,
with filament rigidity $l_{\rm b}/L=0.006$.
Dangling ends have been removed,
and the thickness of each line is proportional to the
energy density, with a minimum thickness so that all
rods are visible (most lines take this minimum value here).
The calibration bar shows what proportion of the deformation
energy in a filament segment is due to stretching or bending.
}
\label{f:eg_den1}
\end{figure}

\begin{figure}
\includegraphics[width=8cm]{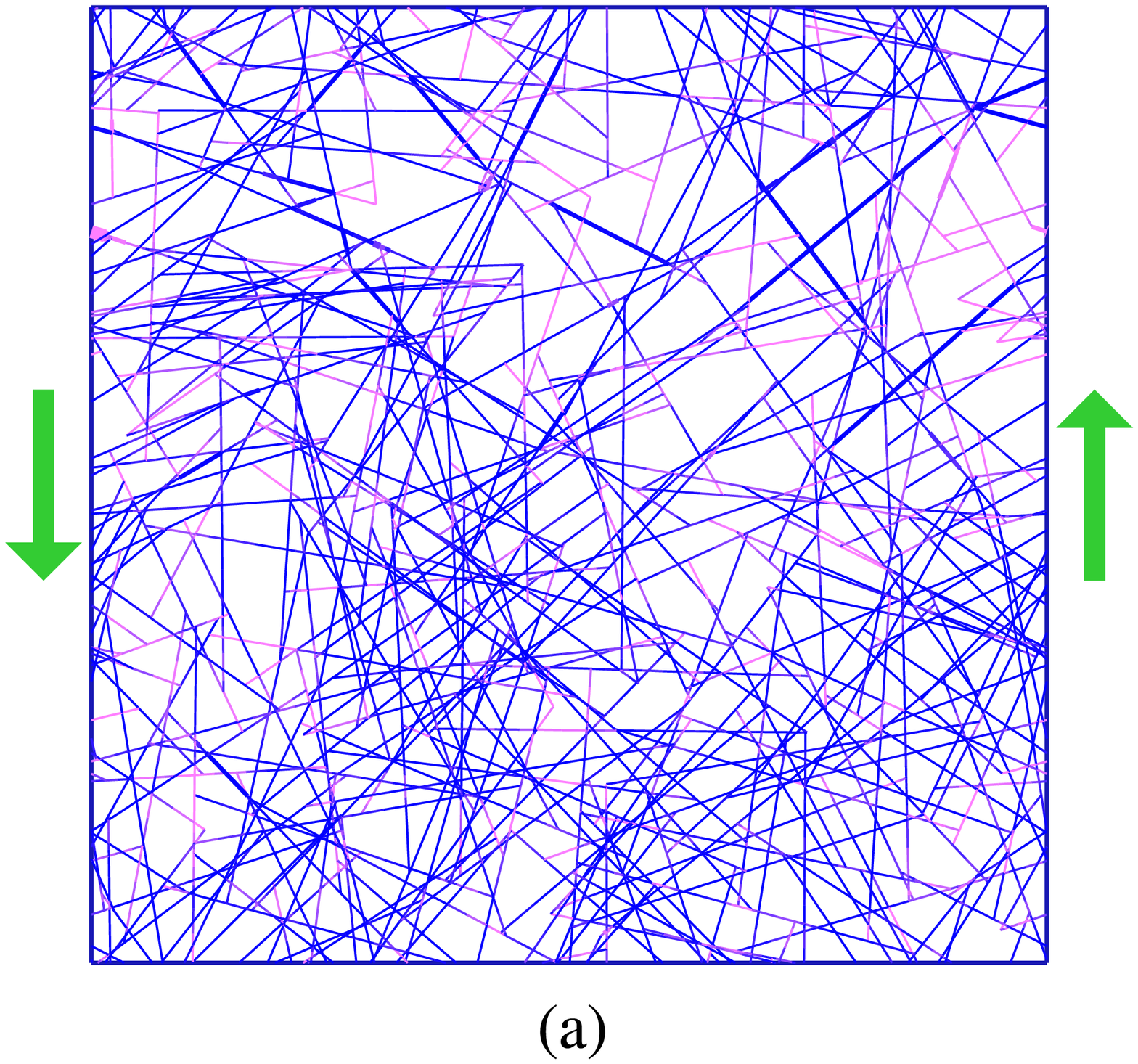}
\includegraphics[width=8cm]{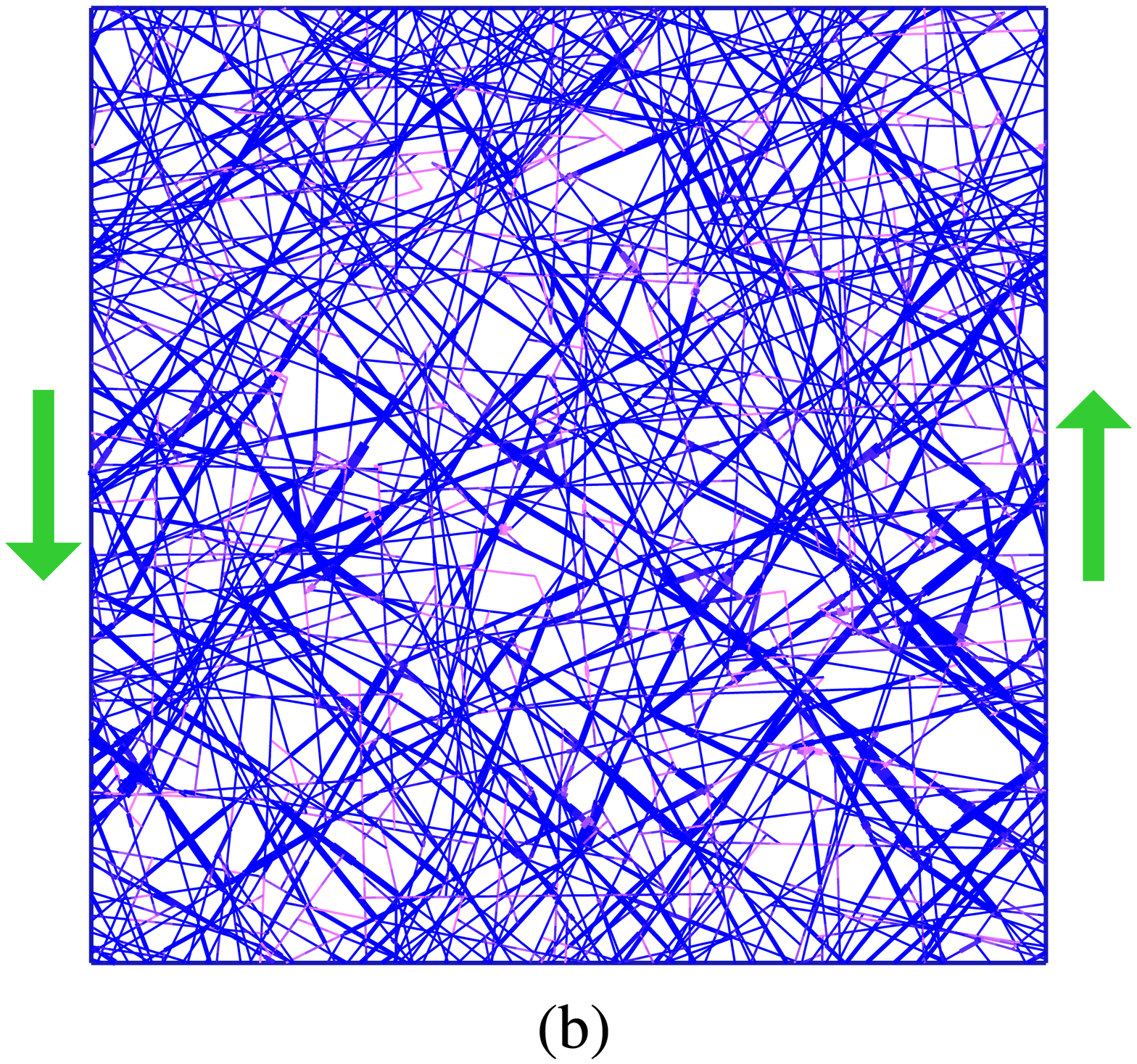}
\caption{{\em Color online}
The same as Fig.~\ref{f:eg_den1} for higher densities
$L/l_{\rm c}\approx29.09$~(a) and
$L/l_{\rm c}\approx46.77$~(b).
For calibration of the colors see Fig.~\ref{f:eg_den1}.
}
\label{f:eg_den2}
\end{figure}

The system Hamiltonian is found by using discrete versions of (\ref{e:H_bend}) and (\ref{e:H_stretch})
which are linearized with respect to filament deflection, ensuring that the macroscopic response is also linear. 
The detailed procedure is described in the Appendix.
It is then minimized to find the network configuration in mechanical equilibrium.
Since entropic effects are ignored: we are formally in the $T\equiv0$ limit.  The filaments are coupled at cross--links, which may exert arbitrary constraint forces but do not apply constraint torques so that the filaments are free to rotate about their crossing points.
We comment on the validity of this assumption in greater detail in Sec.~\ref{s:discuss} below. Specifically, we find that whether or not the physical crosslinks are freely-rotating, the mechanical consequences of such crosslinks is small for dilute networks, provided that the networks are isotropic.
Once the displacements of the filaments under the applied strain of magnitude $\gamma$ have been found, the energy per unit area can be calculated, which within our linear approximation
is equal to $\gamma^{2}/2$ times the shear modulus $G$ or the Young's modulus $Y$, for shear and uniaxial strain, respectively~\cite{LL}. Thus the elastic moduli can be found for a specific network.
The procedure is then repeated for different network realizations and system sizes until a reliable estimate of the modulus is found. See Figs.~\ref{f:eg_den1} and~\ref{f:eg_den2}
for examples of solved networks.

The free parameters in the model are the coefficients $\mu$ and $\kappa$ and the length scales $L$ and $l_{\rm c}$.
We choose to absorb $\kappa$ into a third length scale $l_{\rm b}$ derived from the ratio of $\mu$ and $\kappa$,
\begin{equation}
l_{\rm b}^{2}=\frac{\kappa}{\mu}
\end{equation}
Although the simulations assume a constant angular curvature
between nodes, it is possible to assign $l_{\rm b}$
the physical interpretation of the natural length over which
a free filament bends when differing tangents are imposed
at each end, as seen by a simple Euler minimization of
the Hamiltonian.
Hence we use the subscript $l_{\rm b}$, denoting
`bending length.'
Since $\mu$ now gives the only energy scale in the problem, it scales out, along with one of the length scales (say $L$), and thus we are left with two dimensionless control parameters: the filament rigidity $l_{\rm b}/L$ and
the cross--link density $L/l_{\rm c}$. Note that there is also a fourth length scale,
namely the system size $W$, but all of the results presented below are for sufficiently large systems that the
$W$--dependence has vanished.

We now explore the various deformation regimes of the system beginning with the most fragile, sparse shear--supporting networks that exist just above the rigidity percolation transition.

\section{Rigidity percolation transition}
\label{s:percolation}

For very low cross--link densities $L/l_{\rm c}$ the rods
are either isolated or grouped together into small clusters,
so that there is no connected path between distant
parts of the system and the elastic moduli vanish.
As the density of cross--links is increased,
there is a {\em conductivity} percolation transition 
at $L/l_{\rm c}\approx5.42$ when a connected cluster
of infinite size first appears~\cite{conductivity}.
If there was an energy cost for rotation at cross--links,
an applied shear strain would now induce a stress response
and the elastic moduli would become
non--zero~\cite{wilhelm:03}.
This is also the case when thermal fluctuations
generate stresses along the filaments~\cite{zero_T}.
However, for networks with freely--rotating cross--links
at zero temperature, such as those under consideration here,
the network is able to deform purely by the translation
and rotation of filaments.
Such a {\em floppy} mode costs zero energy and thus
the elastic moduli remain zero.
This continues to be the case until the
{\em rigidity} percolation transition at a higher density
$L/l_{\rm c}\approx5.93$~\cite{latva_kokko,rigid_review,astrom1,astrom2},
when there are sufficient extra constraints that
filaments must bend or stretch and the moduli become
non--zero.

A full description of the network behavior just above
the transition has been given elsewhere~\cite{head:03a},
so here we summarize the results.
Just above the rigidity transition,
both $G$ and $Y$ increase continuously from zero as a power in $L/l_{\rm c}$, with different prefactors but the same exponent~$f$,

\begin{equation}
G,\:Y \sim
\left(
\frac{L}{l_{\rm c}}
-
\left.\frac{L}{l_{\rm c}}\right|_{\rm trans}
\right)^{f}.
\end{equation}

\noindent{}We have found that
$f=3.0\pm0.2$ \cite{head:03,head:03a},
consistent with the value $3.15\pm0.2$ found independently~\cite{wilhelm:03}.
It is also possible to measure geometric properties
of rigid clusters, such as their fractal dimension;
this has been done using the pebble game method,
and found to give exponents that are similar to
that of  central force ({\em i.e.} Hookean spring) percolation
on a lattice~\cite{latva_kokko}.
However, such networks cannot support bending,
whereas we have found that the system Hamiltonian
for our model
is dominated by its bending term near to the transition.
We conclude that our system is 
in a different universality class to central force percolation,
at least as far as the ideas of universality apply to rigidity percolation;
indeed, it casts doubt on the validity of universality for
force percolation as a whole, as discussed below.
Note that this discrepancy cannot be due to any form of
long range correlation in the morphology of the system,
since our random networks are constructed in such a way
as the ensure geometric correlations cannot extend
beyond the length of a single filament.
Note also that although cross--links in our random networks are
connected by filament sections of varying lengths,
thus producing a broad distribution of spring constants
which can also destroy universality according to the
integral expression in~\cite{sahimi_book}, our networks
do {\em not} violate this condition.
This is simply because there is a maximum length
$L$, and hence a minimum spring constant $\mu/L$,
between any two cross--links, ensuring a low--end
cut--off to the distribution of spring constants.
Similar considerations hold for the bending interaction.

Of the exponents that are used to characterize the critical regime,
those measured by the pebble game method
reflect topological or geometric properties of the growing rigid cluster. 
The exponent $f$ is of a different class since it measures the mechanical properties of the fragile solid that appears at the critical point.
Our observation of distinct $f$'s for two systems 
({\em i.e.} our simulations and central force lattices)
that appear to share the same geometric and topological
exponents suggests that, although there are large
universality classes for the topological exponents describing the 
interactions that produce the appropriate number of constraints, the scaling of the shear modulus admits a larger range of relevant perturbations.
We suspect that while rigidity itself is a highly nonlocal property of the network, the modulus depends critically on how stress propagates through particular fragile, low density regions of the rigid cluster. Thus the modulus depends on details of how stress propagates locally through perhaps a few cross-links so that the mechanical characteristics of the filaments and the cross-links become relevant. 

The possibility of experimentally observing the physics associated with the  zero--temperature, rigidity percolation critical point~\cite{head:03a} sensitively depend on the size of the critical region. As with other strictly zero temperature phases transitions, there can be experimental consequences of the critical point physics only if the system can be tuned to pass through the critical region, since the critical point itself cannot be explored. At finite temperature, the network below the rigidity percolation point (but above the connectivity, or scalar percolation point) has a residual static shear modulus generated by entropic tension in the system. One might imagine that one may crudely estimate size of the critical regime around the rigidity percolation transition by comparing the decaying zero-temperature ({\em i.e.} mechanical) modulus of the network above the phase transition $G \sim (L/l_{\rm c} - L/l_{\rm c}|_{\rm trans} )^f$ to the entropic modulus below it, $G_{\rm en}\sim k_{\rm B} T/l_{\rm c}^3$. Unfortunately to make such an estimate one implicitly makes assumptions about the, as yet unknown, cross-over exponents.  Regardless, we speculate that the critical percolation point may indeed have physical implications at room temperature. First, due to the significant stiffness of the filaments,  this residual entropically generated modulus should be small and thus the physics of the zero-temperature critical point may have experimental relevance. Secondly, one may interpret the observed difference between the numerically extracted and calculated scaling exponents $z$ that describes the dependence of $\lambda$ on $l_{\rm c}/l_{\rm b}$ as evidence of corrections to mean-field scaling due to the proximity of the  rigidity percolation critical point. This interpretation is strengthened by the fact that additional data points at smaller values of $l_{\rm c}$ ({\em i.e.} higher cross--link density and thus farther from the rigidity percolation point) but fixed $\lambda$ conform more closely to our mean-field scaling exponent. Thus, it is reasonable to expect that one may observe phenomena associated with rigidity percolation in sparsely cross-linked actin systems.

\section{Elastic regimes}
\label{s:elastic}

The coarse-grained deformation of a material is normally described by the strain field that
is defined at all spatial points. Both the internal state of stress and the density of stored
elastic energy (related by a functional derivative) are then functions of the symmetrized 
deformation tensor. In our model, the underlying microscopic description consists entirely of the
combination of translation, rotation, stretching and bending modes of the filaments. Of these, only the latter
two store elastic energy and thereby generate forces in the material. Thus a complete description of the energetics
of a filament encompassing these two modes must lead to a macroscopic, or continuum elastic description of the material. 

The state of deformation itself, however, is purely a geometric quantity; it can be discussed independently of the energetics associated with the deformation of the filaments themselves. We will characterize the deformation field as {\em affine} if deformation tensor is spatially uniform under uniformly applied strain at the edges of the sample.
Of course, this strict affine limit is never perfectly realized within
our simulations, but, as shown below, there is a broad region
of parameter space in which the deformation field is approximately
affine, in the sense that quantitative measures of the network
response asymtote to their affine predictions.
This entire region shall be called ``affine''.
Since in continuum elastic models the stored elastic energy depends only on the squares (and possibly higher even powers) of the  spatial gradients of the symmetrized deformation tensor, it is clear that uniform strain is a global energy minimum of the system consistent with the uniformly imposed strain at the boundary.  

The spatial homogeneity of the strain field allows one to draw a particularly simple connection between the elastic properties of the individual elements of the network and its collective properties. Because under affine deformation every filament experiences exactly the same deformation, the collective elastic properties of the network can be calculated by determining energy stored in a single filament under the affine deformation and the averaging over all orientations of filaments. Note this calculation constitutes a mean-field description of elasticity.  

If the strain field is purely affine (\emph{i.e.}, is uniformly 
distributed throughout the sample) on
length scales larger than the microscopic length scales 
(\emph{e.g.}, the distance between cross--links), 
then it can be Taylor expanded to
give a locally uniform strain in which all elements
of the strain tensor are constants.
It is then straightforward to see that
the filaments would purely stretch (or compress) and the moduli
would be independent of filament bending coefficient~$\kappa$.
Indeed, it is possible to derive exact expressions for $G$
and $Y$ in this case, as described below.

Conversely, nonaffine deformation on microscopic lengths
arises as a result of filament bending, and hence a dependency on $\kappa$ and well as $\mu$.
This is what we find above the rigidity percolation transition, on increasing network concentration or  molecular weight.
Surprisingly, however, this is not restricted to the neighborhood
of the transition, but constitutes a broad regime of the
available parameter space.
This nonaffine regime is dominated by bending, as can be seen by the fact that $G$ and $Y$ are independent of~$\mu$ below.  It is important to note that continuum elasticity breaks down on length scales over which the deformation field is nonaffine. In addition, the appearance nonaffine deformations invalidates the simple, mean-field calculation of the moduli which assumes that every filament undergoes the deformation. Generically, the moduli in the nonaffine deformation regime will be smaller than their value calculated under the assumption of affine deformation. Nonaffine deformation fields in effect introduce more degrees of freedom since the deformation field is nonuniform. Using those extra degrees of freedom, the system is able to further lower its elastic energy by nonaffine deformations and thereby reduce its modulus. Only upon increasing the number of mutual constraints in the system, can one constrain the system to affine deformations and thereby maximize its modulus.

\subsection{Nonaffine, bending--dominated}
\label{s:bend}

Starting from the most sparse networks just above the rigidity percolation transition, we first encounter the nonaffine 
regime on increasing $L/l_{\rm c}$.
We find empirically that throughout this regime (until the cross-over to affine deformation of one kind or the other -- $AM$ or $AE$) the moduli  of the network are controlled by the bending modes of the filaments.
This nonaffine response can be distinguished from the
scaling regime around the transition, which is also nonaffine,
by the lack of the diverging length scale associated
with a continuous phase transition;
in the simulations, this corresponds to the
independence of $G$ and $Y$ on system size $W$
for relatively small~$W$, as opposed to the increasingly
large $W$ required for convergence close to the transition.

The dependence of $G$ on the system parameters
in this regime can be semi--quantitatively understood
as follows.
Consider what happens when $\kappa\rightarrow0$.
In this limit, filaments can freely bend and only stretching
modes contribute to the dynamic response.
This {\em Hookean} case has already
been investigated by Kellom\"aki {\em et al.}
for the same random network geometries
as considered here~\cite{zero_G}.
They found the striking result that floppy modes
exist for all densities in the linear response,
{\em i.e} $G\equiv Y\equiv 0$.
In these floppy modes, the filaments will bend
at cross--links, but without costing energy since $\kappa\equiv0$.
If $\kappa$ is now continuously increased from zero,
then there should be a range of sufficiently small $\kappa$
in which the angles remain unchanged but
now incur an energy cost according to~(\ref{e:H_bend}),
giving a total energy and hence $G$ that
is proportional to $\kappa$.

To make this idea more specific, suppose that 
when $\kappa\equiv0$ the angles of filament deflection
at cross--links,
$\{\delta\theta\}$, are distributed with zero mean 
and variance $\sigma^{2}_{\delta\theta}$\,.
By assumption, $\sigma^{2}_{\delta\theta}$ can only
depend on the two geometric lengthscales $L$ and $l_{\rm c}$,
or to be more precise on their ratio $L/l_{\rm c}$.
From~(\ref{e:H_bend}),
the energy at each cross--link is

\begin{equation}
\delta{\mathcal H}_{\rm bend}
\sim
\kappa
\left(\frac{\sigma_{\delta\theta}}{l_{\rm c}}\right)^{2}
l_{\rm c}
\label{e:dh_bend}
\end{equation}

\noindent{}The mean number of cross--links per unit area
is $NL/2l_{\rm c}$, where $N$ is the number
of filaments per unit area.
$N$ can be exactly related to $L/l_{\rm c}$ using the
expression derived in~\cite{head:03a};
however, for current purposes it is sufficient to use
the approximate relation
$L/l_{\rm c}\approx(\alpha-1)/(1-2/\alpha)$,
where $\alpha=2L^{2}N/\pi$.
Thus (\ref{e:dh_bend}) summed over the
whole network gives

\begin{equation}
G_{\rm bend}
\sim
\kappa
\frac{\sigma^{2}_{\delta\theta}}{l_{\rm c}^{3}}
\label{e:G_bend}
\end{equation}

\noindent{}Therefore plotting
$GL/\mu=\sigma^{2}_{\delta\theta}Ll^{2}_{\rm b}/l_{\rm c}^{3}$
versus $l_{\rm b}/L$ on log--log axes will give a straight
line of slope~2, as confirmed by the simulations below.
It is also possible to infer the variation of
$\sigma_{\delta\theta}$ on $L/l_{\rm c}$ from either
the simulations or the scaling argument presented below,
but since this is not an easily measurable quantity
experimentally, nothing more will be said about it here.
A bending--dominated response was assumed
in the calculations of Frey {\em et al.}~\cite{frey:98}
and Joly--Duhamel {\em et al.}~\cite{duhamel},
although the 2D and 3D density dependencies
will of course be different.

\subsection{Affine, stretching--dominated}
\label{s:G_affine}
Under an affine strain, the network response consists
purely of stretching modes and it is straightforward to
calculate the corresponding modulus~$G_{\rm affine}$\,.
Consider a rod of length $L$
lying at an angle $\theta$ to the $x$--axis.
Under a shear $\gamma_{\rm xy}$,
this will undergo a relative change in length
$\delta L/L=\gamma_{\rm xy}\sin\theta\cos\theta$,
and therefore an energy cost

\begin{equation}
\delta{\cal H}_{\rm stretch}
=
\frac{1}{2}\mu
L
\gamma_{\rm xy}^{2}
\sin^{2}\theta
\cos^{2}\theta
\label{e:dH_affine}
\end{equation}

\noindent{}The $\sin^{2}\theta\cos^{2}\theta$ factor reduces
to $1/8$ after uniformly averaging over all
$\theta\in(0,\pi)$.
Summing over the network as in the nonaffine case
gives

\begin{eqnarray}
G_{\rm affine}
&=&
\frac{2N\langle\delta{\cal H}_{\rm stretch}
\rangle_{\theta}}{\gamma_{\rm xy}^{2}}
\nonumber\\
&\approx&
\frac{\pi}{16}\frac{\mu}{L}
\left(
\frac{L}{l_{\rm c}}
+2\frac{l_{\rm c}}{L}
-3
\right)
\label{e:G_affine}
\end{eqnarray}

\noindent{}where we have also corrected for dangling ends
by renormalising the rod lengths $L\rightarrow L-2l_{\rm c}$.
In the high---density limit $L/l_{\rm c}\rightarrow\infty$,
(\ref{e:G_affine}) asymtotes to

\begin{equation}
G_{\rm affine}
\sim
\frac{\pi}{16}\frac{\mu}{l_{\rm c}}
\end{equation}

\noindent{}Since the number of rods
per unit area is $N\sim1/(Ll_{\rm c})$, 
then the concentration of protein monomers
of characteristic size $a$ is
$c\sim NL/a\sim(al_{\rm c})^{-1}$ and thus
$G\sim(ac)^{\alpha}$ with $\alpha=1$.
For comparison, theories of thermal 3D systems
predict $\alpha=\frac{7}{5}$ for calculations based
on a tube picture~\cite{hinner:98,isambert:96,morse:98},
$\alpha=\frac{11}{5}$ for affine scaling
relations~\cite{mackintosh:95},
and $\alpha=2$ for the $T=0$ three--dimensional
cellular foam~\cite{satcher96}.

The above calculation can be repeated for
an affine uniaxial strain $\gamma_{\rm yy}$
to give similar expressions for the Young's
modulus~$Y$,
with the $\sin^{2}\theta\cos^{2}\theta$ term in
(\ref{e:dH_affine}) replaced with $\sin^{4}\theta$.
Since this averages to $3/8$, $Y_{\rm affine}$
differs by a factor of 3,

\begin{equation}
Y_{\rm affine}=3G_{\rm affine}
\end{equation}

\noindent{}Hence the Poisson ratio
$\nu=Y/2G-1$
for affinely sheared networks is
$\nu_{\rm affine}=\frac{1}{2}$,
which should be compared to the 3D lattice prediction
$\nu=\frac{1}{3}$~\cite{satcher96}
and the 3D Cayley tree value
$\nu=\frac{1}{4}$~\cite{jones_ball}.

\subsection{Scaling argument and crossover between elastic regimes}
\label{s:scaling}

We now attempt to identify the dominant mode governing the deviation from the affine solution.
The relevant length scales for this mode are derived, which, by comparing to other lengths in the problem,
allow us to estimate when the crossover between stretching--dominated
and bending--dominated regimes should occur. This prediction correctly predicts the qualitative trends
of the deviation from affinity with the lengths $L$, $l_{\rm c}$ and $l_{\rm b}$. However, it is not as successful quantitatively as an empirical scaling law described in Sec.~\ref{s:moduli}. Nonetheless we believe it contains the essential physics and therefore warrants a full description.

To proceed, we note that the stretching--only solution presented above assumes that the stress is uniform
along a filament until reaching the dangling end. It is more realistic to suppose that it vanishes smoothly.
If the rod is very long, far from the ends and near the center of the rod it is
stretched/compressed according to the macroscopic strain $\gamma_0$. We assume that this
decreases toward zero near the end, over a length $l_\parallel$, so that the reduction in stretch/compression energy is of order $\mu\gamma_{0}^2l_\parallel$.  The amplitude of the displacement along this segment, which is located near the ends of the rod is of order $d\sim\gamma_{0} l_\parallel$. This deformation, however, clearly comes at the price of deformations of surrounding 
filaments, which we assume to be primarily bending in nature (the dominant constraints on this rod will be due
 to filaments crossing at a large angle). The typical amplitude of the induced curvature is of order
$d/l_\perp^2$, where $l_\perp$ characterizes the range over which
the curved region of the crossing filaments extends. This represents what
can be thought of as a bending correlation length, and it will be, 
in general, different from $l_\parallel$. The latter can also be thought of as a correlation length, specifically for the strain variations near free ends. We determine these lengths self-consistently, as follows.

The corresponding total elastic energy contribution due to these coupled
deformations is of order
\begin{equation}
\Delta E_1=-\mu\gamma_{0}^2l_\parallel
+\kappa\left(\frac{\gamma_{0}
l_\parallel}{l_\perp^2}\right)^2l_\perp\frac{l_\parallel}{l_{\rm c}},
\end{equation}
where the final ratio of $l_\parallel$ to $l_{\rm c}$ gives the typical number of constraining
rods crossing this region of the filament in question. In simple physical terms, the rod can reduce its total elastic energy by having the strain near the free ends deviate from the otherwise affine, imposed strain field. In doing so, it results in a bending of other filaments to which it is coupled. From this, we expect that the range of the
typical longitudinal displacement $l_\parallel$ and transverse displacement $l_\perp$ are
related by
\begin{equation}
l_\perp^3\sim l_{\rm b}^2l_\parallel^2/l_{\rm c}.
\end{equation}

Of course, the bending of the other filaments will only occur 
because of constraints on them. Otherwise, they would simply translate in space.
We assume that the transverse constraints on these bent filaments to be primarily due to
compression/stretch of the rods which are linked to them. These distortions will be governed by the same physics as described above. In particular, the length scale of the corresponding deformations is of order $l_\parallel$,
and they have a typical amplitude of $d$. Thus, the combined curvature and stretch energy is of order

\begin{equation}
\Delta E_2=\mu\gamma_{0}^2l_\parallel\frac{l_\perp}{l_{\rm c}}
+\kappa\left(\frac{\gamma_{0}
l_\parallel}{l_\perp^2}\right)^2l_\perp,
\end{equation}
where, in a similar way to the case above, $l_\perp/l_{\rm c}$ determines the typical number of filaments constraining the bent one we focus on here. This determines another relationship between the optimal bending and stretch correlation lengths, which can be written

\begin{equation}
l_\perp^4\sim l_{\rm b}^2l_\parallel l_{\rm c}.
\end{equation}

\noindent{}Thus, the longitudinal strains of the filaments decay to zero over a length of order

\begin{equation}
\label{ell-parallel}
l_\parallel\sim l_{\rm c} \left(\frac{l_{\rm c}}{l_{\rm b}}\right)^{2/5},
\end{equation}

\noindent{}while the resulting bending of filaments extends over a distance of order

\begin{equation}
\label{ell-perp}
l_\perp\sim l_{\rm c} \left( \frac{l_{\rm b}}{l_{\rm c}}\right)^{2/5}.
\end{equation}
The physical implications of Eqs.~(\ref{ell-parallel}) and (\ref{ell-perp}) is that a length of each filament of order $l_\parallel$ experiences nonaffine deformation and this nonaffine deformation causes changes in the local strain field over a zone extending a perpendicular distance $l_\perp$ from the ends of that filament. Thus when $l_\parallel$ becomes comparable to the length of the filament, $L$, the network should deform in a nonaffine manner. We will refer to this length along the filament contour over which one expects to find nonaffine deformation as $\lambda$.

These results make sense, as increased bending rigidity can be expected to 
increase the bending correlation length $l_\perp$, while decreasing the longitudinal correlation 
length $l_\parallel$ because of the stiff constraints provided by the cross--links. Both lengths, of 
course, tend to increase with decreasing concentration of cross--links, \emph{i.e.}, with increasing $l_{\rm c}$.
This scaling analysis assumes, however, that $l_{\rm c} < l_{\parallel,\perp}$. Furthermore, we expect that $l_{\rm b}<l_{\rm c}$ in general. This is because the bending stiffness of a rod $\kappa\sim {\rm Y}_{f} r^4$ increases with the fourth power of its radius $r$, while $\mu\sim {\rm Y}_{f} r^2$ increases with the square of the radius, where ${\rm Y}_{f}$ is the Young's modulus of the filament. Thus, $l_{\rm b}$ is expected to be of order the rod diameter, which must be smaller than the distance between cross-links, especially considering the the small volume fractions $\phi$ of less than 1\% in many cases. For rods of radius $r$ in three dimensions, we expect that $l_{\rm c}\sim a/\sqrt{\phi}$. Thus, $l_{\rm b}/l_{\rm c}$ may be in the range of 0.01 to 0.1.
This means that we expect that $l_\parallel>l_\perp$, although both of these 
lengths are of order $l_{\rm c}$. Thus, the natural dimensionless variable
determining the degree of affinity of the strain is the ratio of the filament length to $l_\parallel$, the
larger of the two lengths that characterize the range of nonaffine deformation; \emph{i.e.} when the
filaments are very long compared with the effectively stress-free ends, then most of the rod segments
experience a stretch/compression deformation determined by their orientation and the macroscopic strain.

It is possible, however, that $l_\perp$ above may become smaller than $l_{\rm c}$, especially for 
either very flexible rods or for low concentrations. This is unphysical, and we expect the bending of 
constraining filaments above to extend only over a length of order $l_{\rm c}$ when $l_{\rm b}/l_{\rm c}$ becomes 
very small. This results in a different scaling of $l_\parallel$, given by

\begin{equation}
l_\parallel\sim l_{\rm c}^2/l_{\rm b}.
\end{equation}

\noindent{}Although we see no evidence for this scaling, it may become valid for small enough $l_{\rm b}/l_{\rm c}$.

\subsection{Numerical results for elastic moduli}
\label{s:moduli}

\begin{figure}[t] \centering
\includegraphics[width=8cm]{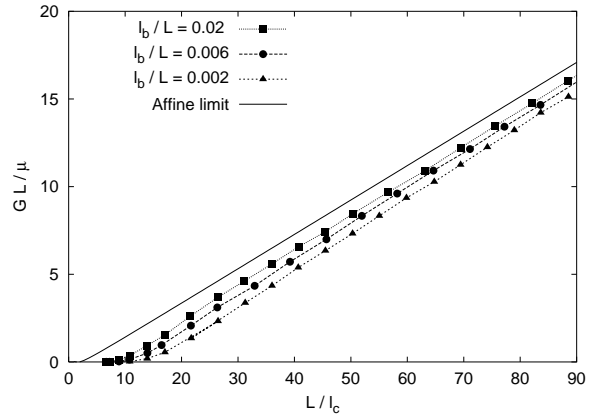}
\caption{The normalized shear modulus $GL/\mu$
versus the cross--link density $L/l_{\rm c}$ for three
different bending lengths $l_{\rm b}/L$.
The solid line gives the affine solution (\ref{e:G_affine})
and the dashed lines adjoining the data points are
to guide the eye.
Here and throughout, errors are no larger than the symbols.}
\label{f:G_with_den}
\end{figure}

\begin{figure}\centering
\includegraphics[width=8cm]{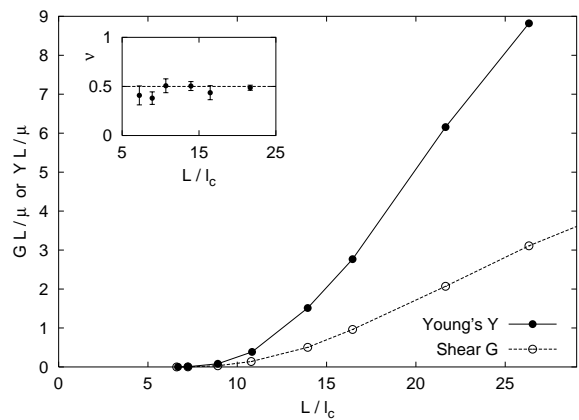}
\caption{The shear and Young's moduli $G$ and $Y$
for $l_{\rm b}/L=0.006$ against $L/l_{\rm c}$.
The interconnecting lines are to guide the eye.
{\em (Inset)}~The Poisson ratio $\nu=Y/2G-1$ against
$L/l_{\rm c}$ for the same $l_{\rm b}/L$.
}
\label{f:youngs}
\end{figure}

We now summarize our numerical results starting first at lowest filament densities. Away from the rigidity transition, the shear modulus $G$ continues to increase monotonically with the cross-link density $L/l_{\rm c}$ at a rate that only weakly depends on the filament rigidity $l_{\rm b}/L$, as shown in Fig.~\ref{f:G_with_den}.
Indeed, for high densities $G$ approaches the affine
prediction $G_{\rm affine}$ which, due to the
absence of bending modes under affine strain as discussed at the beginning of 
Sec.~\ref{s:elastic}, is independent of $l_{\rm b}$. The filament rigidity $l_{\rm b}$ does, however,
influence the crossover to the affine solution as will be discussed below.
The Young's modulus $Y$ and the Poisson ratio $\nu=Y/2G-1$ for a range of densities are shown
in Fig.~\ref{f:youngs}. It is apparent that $\nu$ remains close to the affine prediction $\nu_{\rm affine}=\frac{1}{2}$, except possibly near the transition where it
takes the value $0.35\pm0.1$~\cite{head:03a} (here as elsewhere, quoted errors are single standard deviations).
Note that, in two dimensions, area--preserving deformations have $\nu=1$, so $\nu=0.5$
does not imply incompressibility. For comparison, the thermodynamically stable range is $-1\leq\nu\leq1$~\cite{LL}.
The robustness the affine prediction of the Poisson ratio even deep in the $NA$ regime is somewhat surprising and is 
not accounted for in our arguments.  

\begin{figure}\centering
\includegraphics[width=8cm]{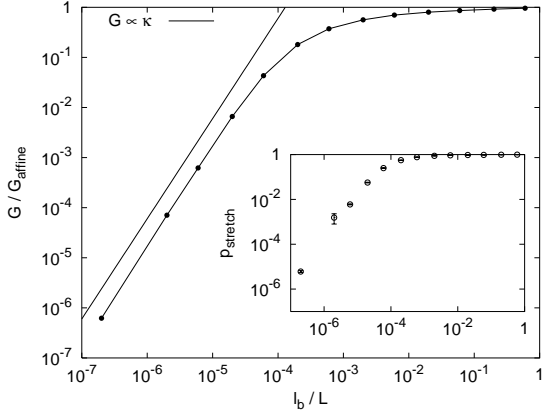}
\caption{Shear modulus $G$ versus filament rigidity
$l_{\rm b}/L$ for $L/l_{\rm c}\approx29.09$,
where $G$ has been scaled to the affine prediction
for this density.
The straight line corresponds to the bending--dominated
regime with $G\propto\kappa$, which gives a line of
slope 2 when plotted on these axes.
{\em (Inset)}~The proportion of stretching energy to the total
energy for the same networks, plotted against
the same horizontal axis $l_{\rm b}/L$.
}
\label{f:unscaled}
\end{figure}

Varying the ratio $l_{\rm b}/L$ over many orders of magnitude
at fixed $L/l_{\rm c}$
reveals a new regime in which $G\propto\kappa$,
rather than $G\sim G_{\rm affine}\propto\mu$
as in the affine regime described above.
An example is given in  Fig.~\ref{f:unscaled},
where it can be seen that $G\propto l_{\rm b}^{2}\propto\kappa$,
suggesting that this regime is
dominated by {\em bending} modes,
a claim that is supported by the theoretical considerations presented in Sec.~\ref{s:bend} and the work of Frey {\em et al.}~\cite{frey:98} and Joly--Duhamel {\em et al.}~\cite{duhamel}.
We also confirm in the inset to this figure that the regime for which $G\approx G_{\rm affine}$ is dominated by stretching modes, and that this new regime with $G\propto\kappa$
is dominated by bending modes, as expected.

\begin{figure}
\includegraphics[width=8cm]{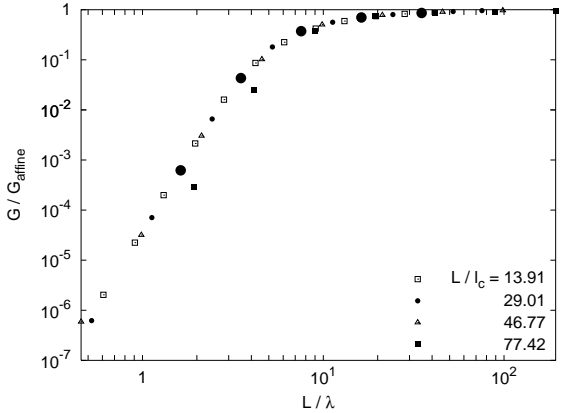}
\caption{$G/G_{\rm affine}$
versus $L/\lambda$ with
$\lambda=\sqrt[3]{l_{\rm c}^{4}/l_{\rm b}}$
for different densities $L/l_{\rm c}$, showing good collapse
except for the highest density considered.
The enlarged points for $L/l_{\rm c}\approx29.09$
correspond to the same parameters
as in Fig.~\ref{f:aff_with_r}.}
\label{f:lambda1}
\end{figure}

\begin{figure}
\includegraphics[width=8cm]{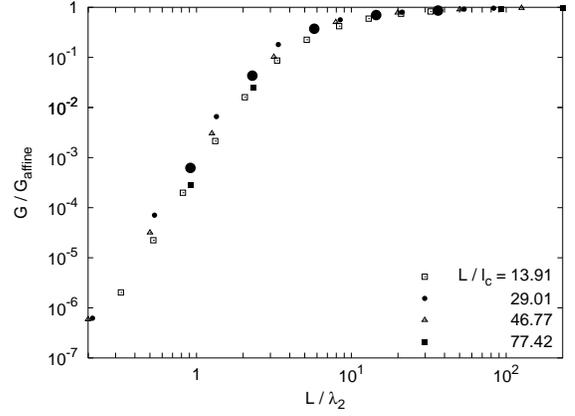}
\caption{The same data as Fig.~\ref{f:lambda1}
plotted against
$L/\lambda_{2}$ with
$\lambda_{2}=\sqrt[5]{l_{\rm c}^{7}/l_{\rm b}^{2}}$,
as predicted by the scaling argument in the text,
showing slight but consistent deviations from
collapse for this range of $L/l_{\rm c}$.
}
\label{f:lambda2}
\end{figure}

The crossover between the two regimes can be
quantified by the introduction of a new length $\lambda$,
being a combination of $l_{\rm b}$ and $l_{\rm c}$
characterized by an exponent~$z$,

\begin{equation}
\label{lambda-exp}
\lambda= l_{\rm c} \left(\frac{l_{\rm c}}{l_{\rm b}}\right)^z.
\end{equation}

\noindent{}The ratio $L/\lambda$ can then be used to
ascertain which regime the network is in,
in the sense that $\lambda\ll L$ corresponds to
the affine regime,
and $\lambda\gg L$ corresponds to
the nonaffine regime.
Note that this is only possible outside the neighborhood
of the rigidity transition.
For densities in the approximate range $13<L/l_{\rm c}<47$,
a very good data collapse can be found by using
$\lambda_{1}=\sqrt[3]{l_{\rm c}^{4}/l_{\rm b}}$
or $z=\frac{1}{3}$, as demonstrated in Fig.~\ref{f:lambda1}.
This empirical relation has already been published~\cite{head:03}.
However, as clearly evident from the figure,
it appears to fail for the small number of very high
density points that we have now been able to attain.
Conversely, the scaling argument of Sec.~\ref{s:scaling}
generates the relevant length scale
$\lambda_{2}=l_{\parallel}=\sqrt[5]{l_{\rm c}^{7}/l_{\rm b}^{2}}$
or $z=\frac{2}{5}$.
Although the data does not collapse for this second form,
as evident from Fig.~\ref{f:lambda2},
it appears to improve the overall collapse for larger
densities ({\em i.e.} further from the rigidity transition point).
A possible explanation for this is that $\lambda_{2}$ is
the correct asymptotic length, but does not apply for
networks of intermediate density, where the empirical
form $\lambda_{1}$ is much more successful,
perhaps due to corrections to scaling from the transition
point as discussed in Sec.~\ref{s:percolation}.
(This is also the relevant range for biological applications).
However, our current computational resources cannot
go to higher densities, and so we must leave this question
to be resolved at a later date, by either improved theory or
increased processor speeds.

Although the goal of this paper is to characterize the behavior of semiflexible polymer networks over
the whole of the parameter space, it is nonetheless instructive to also consider parameters corresponding
to physiological actin networks. The lengths $L$, $l_{\rm c}$ and $l_{\rm b}$ for F--actin
in physiological conditions can be approximated as follows. The distance between cross--links has been quoted
as $l_{\rm c}\approx0.1\mu m$~\cite{frey:98}, which for filament lengths $L\approx2\mu m$
gives a cross--link density $L/l_{\rm c}\approx20$. As already argued in Sec.~\ref{s:scaling},
we can estimate $l_{\rm b}$ by regarding the filament as a solid elastic cylinder with radius $r$, in which case $l_{\rm b}\sim r\approx 10nm$.  Thus we find that $l_{\rm b}/L\sim r/L\sim10^{-3}$, which gives $L/\lambda\approx5$.
Looking at Fig.~\ref{f:lambda1}, this suggests that cytoskeletal networks are in the crossover region.
Similarly, $L/\lambda_2\approx4$ leading to the
same basic conclusion.

\subsection{Spatial correlations}

A further way of probing the degree of affinity of
a network is to consider a suitable spatial
correlation function.
Whatever quantity is chosen, it is clear that there
can be no fluctuations if the strain is purely affine.
This is not the case with nonaffine strains, which
will induce localized bending modes that couple to
the local geometry of the network and thus may
fluctuate from one part of the network to the next.
Thus correlations between fluctuations should have
a longer range when the deformation is more nonaffine,
qualitatively speaking.
The two--point correlation function between 
spatially--varying quantities $A({\bf x})$ and
$B({\bf x})$ can be generally defined as

\begin{equation}
C_{AB}(r)=
\langle
A({\bf x})B({\bf x}+r\hat{\bf n})
\rangle
-
\langle A({\bf x})\rangle
\langle B({\bf x})\rangle
\end{equation}

\begin{figure}\centering
\includegraphics[height=8cm,angle=270]{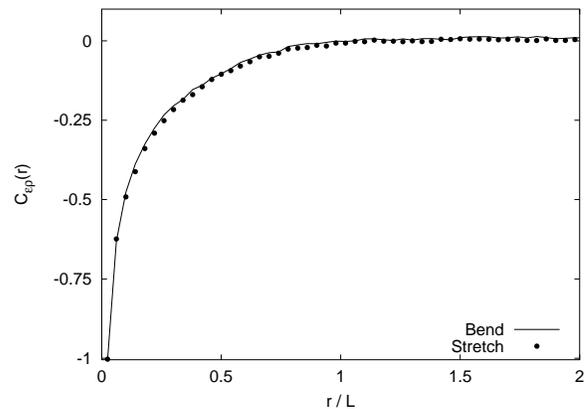}
\caption{The correlation function $C_{\varepsilon\rho}(r)$
between local mass density $\rho$ and energy density
$\varepsilon$, where $\varepsilon$ is restricted to either
stretching or bending energy as shown and
$\rho$ is the length of filaments within a radius
$L/4$ of the network point.
The cross--link density $L/l_{\rm c}\approx21.48$,
$l_{\rm b}/L=0.006$ and
both lines have been normalized so that
$|C_{\varepsilon\rho}(r=0)|=1$.
}
\label{f:E_rho}
\end{figure}

\noindent{}where the angled brackets denote averaging
over all network nodes ${\bf x}$ and direction unit vectors
$\hat{\bf n}$.
Fig.~\ref{f:E_rho} shows an example of $C_{\varepsilon\rho}$,
where $\rho$ is the local mass density of filaments
and $\varepsilon$ is the energy per unit filament length,
restricted to either stretching or bending energy
as shown in the key.
There is a clear anti--correlation between density and both
forms of energy at short separations,
showing that the magnitude of deformation is
heterogeneously distributed throughout the network,
being greater in regions of low mass density and
smaller in regions of high mass density.
Qualitatively, the network concentrates the largest
deformations into low mass density regions, thus
reducing the macroscopic energy cost.
This effects both bending and stretching modes equally:
there is no increased likelihood of one mode over the other
for regions of given density,
as demonstrated by the collapse of
$C_{\varepsilon\rho}$ for both energy types collapse
after normalization, also given in this figure.

\begin{figure}
\includegraphics[height=8cm,angle=270]{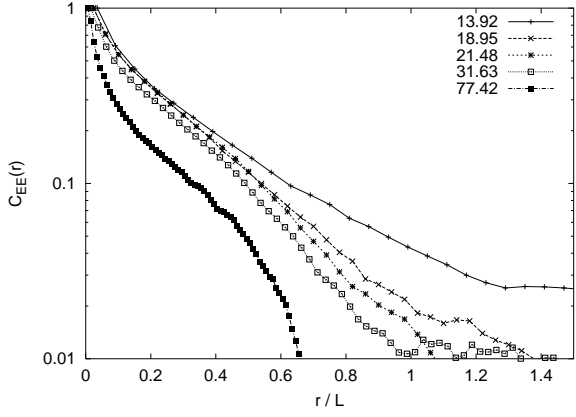}
\caption{Autocorrelation of the combined (stretching plus
bending) energy density $E$ for $l_{b}/L=0.006$
and the $L/l_{\rm c}$ given in the key.
The system sizes were
$W=15L$ ($L/l_{\rm c}=13.92$),
$10L$ (18.95 and 21.48),
$6\frac{1}{4}L$ (31.63) and
$2\frac{1}{2}L$ (77.42).
}
\label{f:E_E}
\end{figure}

The sizes of locally--correlated regions can be
inferred from the decay of a suitable autocorrelation function,
such as the combined energy $E$ (stretching plus bending)
per unit length.
$C_{EE}(r)$ is plotted in Fig.~\ref{f:E_E} for different
density networks.
The trend is for $C_{EE}(r)$ to decay more slowly with $r$
for lower $L/l_{\rm c}$ at fixed $l_{\rm b}/L$,
suggesting larger `pockets' of non--uniform deformation
for lower network densities, presumably becoming
infinitely large at the transition, where the correlation
length diverges algebraically with the known
exponent $\nu\approx1.17\pm0.02$~\cite{latva_kokko}.
We have been unable to extract a meaningful length
scale from our $C_{EE}(r)$ data and hence are unable
to confirm the value of this exponent.

\subsection{Measures of affinity}
\label{s:affinity}

Intuitively, the degree to which the network deformation is
or is not affine depends on the length scale on which we look.
For length scales comparable to the system size,
the deformation must appear affine since we are imposing
an affine strain at the periodic boundaries.
Only on some smaller length scale might deviations
from affinity be observed.
If the deformation field is nonaffine on length
scales corresponding to the microscopic lengths $L$,
$l_{\rm c}$ or $l_{\rm b}$,
then the filaments will `feel' a locally non--uniform strain field
and the assumptions leading to the prediction of
$G_{\rm affine}$ will break down.

To quantify the degree of affinity at a given length scale, 
consider the infinitesimal change in angle
under an imposed shear strain between two network nodes
separated by a distance~$r$.
Denote this angle $\theta$, and its corresponding
affine prediction $\theta_{\rm affine}$.
Then a suitable measure of deviation from
affinity on length scales $r$ is

\begin{equation}
\langle\Delta\theta^{2}(r)\rangle
=
\langle
(\theta-\theta_{\rm affine})^{2}
\rangle
\label{e:aff_with_r}
\end{equation}

\begin{figure}[h]\centering
\includegraphics[width=8cm]{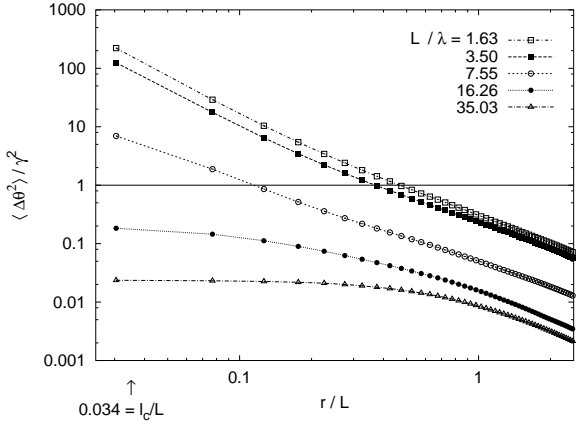}
\caption{Plot of the affinity measure
$\langle\Delta\theta^{2}(r)\rangle$
normalized to the magnitude of the imposed strain $\gamma$
against distance~$r/L$, for different $l_{\rm b}/L$.
The value of $r$ corresponding to the mean distance
between cross--links $l_{\rm c}$ is also indicated,
as is the solid line
$\frac{1}{\gamma^{2}}\langle\Delta\theta^{2}(r)\rangle=1$,
which separates affine from nonaffine networks
to with an order of magnitude
(the actual crossover regime is somewhat broad).
In all cases, $L/l_{\rm c}\approx29.1$
and the system size was $W=\frac{15}{2}L$.
}
\label{f:aff_with_r}
\end{figure}

\noindent{}where the angled brackets denote averaging
over both network points and different network realizations.
An example of $\langle\Delta\theta^{2}(r)\rangle$
is given in Fig.~\ref{f:aff_with_r},
and clearly shows that it monotonically decays
with distance, as intuitively expected.
Also, the deviation from affinity is uniformly higher
for lower $l_{\rm b}/L$ at the same $L/l_{\rm c}$,
in accord with the greater deviation of $G$ from
$G_{\rm affine}$ observed above.

Although $\langle\Delta\theta^{2}(r)\rangle$
decreases monotonically with~$r$,
the decay is slow, almost power--law like over the ranges
given.
This suggests that there is no single `affinity length scale'
above which the deformation looks affine, and below which
it does not.
However, we can read off the degree of affinity
at the cross--link length scale $r=l_{\rm c}$,
which (after normalising to the strain $\gamma$)
should by $\ll1$ for an affine deformation,
and $\gg1$ for a nonaffine one.
This is evident when comparing Fig.~\ref{f:aff_with_r}
to the $G/G_{\rm affine}$ for the same systems in
Fig.~\ref{f:lambda1};
$\frac{1}{\gamma^{2}}\langle\Delta\theta^{2}(l_{\rm c})\rangle\ll1$
does indeed correspond to $G\approx G_{\rm affine}$, and
$G\approx G_{\rm bend}$ for
$\frac{1}{\gamma^{2}}\langle\Delta\theta^{2}(l_{\rm c})\rangle\gg1$.

The monotonically increasing deviation from affinity
with decreasing $L/\lambda$ can also be seen using
an independent affinity measure, as used by Langer
and Liu~\cite{langer_liu}.
Consider the displacements $\{{\bf\delta x}_i\}$
of each node $i$ after the strain has been applied,
relative to their unstrained positions.
Each of these has a corresponding affine
prediction ${\bf\delta x}_{i}^{\rm affine}$ that
can be simply computed given the node's original
position and the type of strain applied.
Then a scalar measure of the global deviation
from affinity is the root mean square of the difference
between the measured displacements and their
affine values, {\em i.e.}

\begin{equation}
\affmeas^{2}
=
\langle
(\delta{\bf x}_{i}-\delta{\bf x}_{i}^{\rm affine})^{2}
\rangle
\end{equation}

\begin{figure}[h]\centering
\includegraphics[width=8cm]{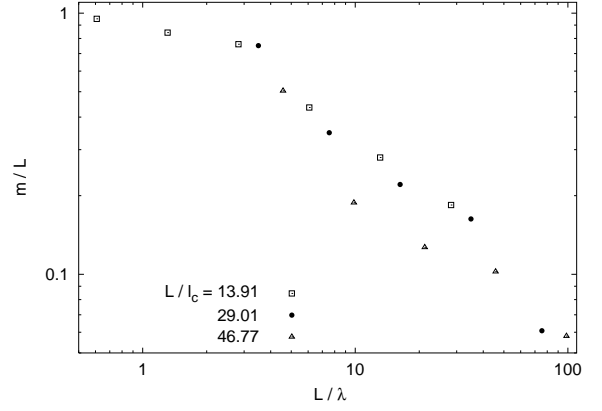}
\caption{The root mean square deviation of
node displacements from their affine prediction, $\affmeas/L$,
plotted against the same $L/\lambda_{1}$ as in
Fig.~\ref{f:lambda1}.
Symbol sizes are larger than errors, so the apparent
scatter is real.
}
\label{f:aff_meas}
\end{figure}

\noindent{}where the angled brackets denote
averaging over all nodes~$i$.
In Fig.~\ref{f:aff_meas} we plot $\affmeas/L$
against $L/\lambda$, and observe the expected
monotonic increase of the deviation from affinity
with decreasing $L/\lambda$.
However, the data for different $L/l_{\rm c}$ do
{\em not} collapse. 
The problem is that, unlike in Figs.~\ref{f:lambda1}
and~\ref{f:lambda2},
it is not obvious how $\affmeas$ should be normalised
to give a dimensionless quantity; we have tried
using the lengthscales $L$, $l_{\rm b}$, $l_{\rm c}$
and $\lambda$, but none of these generate good collapse.
It is likely that some combination of these lengths
{\em will} collapse the data,
but we have been unable to
find it empirically, and we have no theoretical prediction
for this affinity measure.

\section{Thermal effects}
\label{s:thermal}

In the above, we have considered a mechanical, purely athermal model
of networks governed by two microscopic energies: (i) the bending of
semiflexible filaments, and (ii) the longitudinal compliance 
of these filaments that describes their response to compression and
stretching forces. We have already discussed the role of temperature in terms of the formation of a solid via the rigidity percolation transition. Now, we examine the transition between the $AM$ and $AE$ regimes.

For a homogeneous filament of Young's modulus ${\rm Y}_f$,
the corresponding single-filament parameters $\kappa$ (the 
bending rigidity) and $\mu$ (the one-dimensional compression/stretch 
modulus) are determined by ${\rm Y}_f$ and geometric factors:
$\kappa\sim {\rm Y}_f a^4$, and $\mu\sim {\rm Y}_f a^2$, where $2a$ is the filament 
diameter. From here on, we refer to the later (mechanical) modulus 
as $\mu_{M}$. 

At finite temperature, however, there will be transverse fluctuations
of the filaments that give rise to an additional longitudinal compliance. Physically, this compliance comes from the ability to pull-out the thermal fluctuations of the filament, even without any stretching of the filament backbone. The corresponding linear modulus for a filament segment of length $l$ is \cite{mackintosh:95}
\begin{equation}
\mu_{T}=\frac{\kappa\ell_p}{l^3}\sim\frac{a^2\ell_p}{l^3}\mu_M ,
\end{equation}
where $\ell_p=\kappa/(kT)$ is the persistence length. The
full compliance of such a segment is then 
$l/\mu=l/\mu_M+l/\mu_T$, corresponding to an effective 
linear modulus of
\begin{equation}
\mu=\frac{\mu_M\mu_T}{\mu_M+\mu_T}.
\end{equation}
Thus, the thermal compliance dominates for lengths larger than $\sqrt[3]{a^2\ell_p}$, while the segment behaves for all practical purposes as a rigid rod with linear modulus $\mu_M$ for
lengths smaller than $\sqrt[3]{a^2\ell_p}$.

Thus, there appear to be two distinct regimes within the 
affine regime: for higher concentrations, specifically for $l_{\rm c}\alt\sqrt[3]{a^2\ell_p}$, the longitudinal compliance of
the filaments is governed by the mechanical compression/stretching 
of filament segments and the modulus is given by 
Eq.\ (\ref{e:G_affine});  this is the $AM$ regime. 
At lower concentrations, specifically for 
$l_{\rm c}\agt\sqrt[3]{a^2\ell_p}$, the single-filament compliance is
dominated by thermal fluctuations and 
\begin{equation}
G\sim\frac{\pi\kappa\ell_p}{16l_{\rm c}^4}.\label{e:G_affine_th}
\end{equation}
This is the $AE$ regime.
For the networks under discussion, which are described by a single variable $l_{\rm c}$ that both 
represents the spacing of filaments and distance between constraints
or cross--links, the boundary between these two affine regimes is
simply determined by concentration, which is naturally measured as $1/l_{\rm c}$. For actin, we estimate the characteristic length $\sqrt[3]{a^2\ell_p}$ to be of order 100 nm. Thus, only when the distance between cross--links is less than a distance of order 100 nm will the bulk response of the network depend on the purely mechanical extension of actin filaments. 

Experimentally, these two affine regimes should be distinguished by 
their scaling dependencies of the linear shear moduli $G$ 
on various parameters. A clearer, qualitative distinction, however, should be seen in their nonlinear behavior. Specifically, in the thermal regime, the maximum strain (either at which the 
network yields or first exhibits non-linear behavior) is
expected to decrease with increasing concentration of polymer
or cross--links \cite{mackintosh:95}. This is because of
the limited extent of thermal compliance,
which decreases for shorter filament segments
that appear more straight. This is in contrast with mechanical 
networks where the nonlinearities are governed by geometry
(\emph{e.g.}, connectivity and orientation of filaments). This
would suggest a concentration-independent maximum or characteristic
strain, as is seen in some colloidal gels \cite{Weitz}. 

Moreover, the actual form of the force-extension relation for a semiflexible polymer in the limit
of segment lengths $l\ll\ell_p$ takes on a universal form, depending only on a characteristic extension $l^2/\ell_p$ and characteristic force $\pi^2\kappa/l^2$. This force-extension 
relation predicts a universal strain stiffening of semiflexible gels in the affine regime, in contrast with a nonlinear modulus in the mechanical affine regime that will be much more dependent in the type of filament \cite{WorkWithPaul}.


\section{Implications and discussion}
\label{s:discuss}

Starting with solutions having no (zero-frequency) elastic behavior, as the filament concentration $c_f$, molecular 
weight (filament length) $L$, or density of cross--links is increased, there is a point where macroscopic elastic behavior is first observed. This is the rigidity percolation transition, and it occurs for a fixed value 
of $L/l_{\rm c}$, where $1/l_{\rm c}$ is a particularly convenient measure of filament concentration, as it represents 
the line density (length per volume) in our 2-d system, apart from a factor of order unity. 

We have shown that there are three distinct elastic behaviors of semiflexible networks above the rigidity transition: (i) when either the molecular weight (filament length) or concentration is low, a nonaffine regime is expected in which the modulus is determined at a microscopic level by filament bending
(transverse compliance)~\cite{frey:98,satcher96};
(ii) as either the molecular weight or concentration increases,
a crossover is expected to an elastic regime in which the deformations are affine (uniform strain) and in which the elastic response at the large scale is governed by the thermal/entropic longitudinal compliance of filament segments; (iii) at still higher concentrations or cross--link densities, this single-filament compliance becomes dominated by the mechanical compliance of bare filament stretching and compression. 

The crossover between (i) and (ii) is given by a fixed  value of $L/\lambda$ of order 10, where $\lambda$ is a microscopic length characterizing the range of nonaffine deformation along a filament backbone. We expect this length to be of order $l_{\rm c}$, the distance between cross--links. But, it should also depend on the filament stiffness through the length $l_{\rm b}=\sqrt{\kappa/\mu}$. In fact, it can only depend on the two lengths $l_{\rm b}$ and $l_{\rm c}$ in our networks. Thus, we expect that $\lambda=l_{\rm c}\left(l_{\rm c}/l_{\rm b}\right)^z$. 
We have presented a scaling argument that shows this for $z=2/5$, while we find empirically that 
$z\simeq 1/3$ for biologically relevant densities. The boundary between nonaffine and affine regimes is thus given by $L\sim\lambda$.
In the mechanically-dominated regime, $l_{\rm b}=\sqrt{\kappa/\mu_M}$, while in the thermal regime, 
\begin{equation}
l_{\rm b}=\sqrt{\kappa/\mu_T}\sim\sqrt{\frac{l_{\rm c}^3\kappa}{a^2\ell_p\mu_M}}.
\end{equation}
Thus, in the mechanical regime, the boundary is given by $L\sim l_{\rm c}^{1+z}$, while in the thermal regime, the boundary is given by $L\sim l_{\rm c}^{1-z/2}$.  In either case, we see that this crossover has a different functional
dependence on concentration, demonstrating once again that the physics of
this crossover is distinct from the rigidity percolation transition.
We show a sketch of the expected diagram of the various regimes
depending on $L$ and $c_f$ in Fig.\ \ref{f:PD}. The boundary between mechanically-dominated and thermal regimes is simply given by $l_{\rm c}\sim\sqrt[3]{a^2\ell_p}$, as we have noted above. 

We can make several additional observations concerning the behavior of real networks, based on our simple model. First, we note the strong dependence of the shear modulus on the cross--link density, as illustrated in Eq.\ \ref{e:G_affine_th} and already noted for 3D affine networks in Ref.\ \cite{mackintosh:95}. In fact, in the 2D networks presented here, we have made no distinction between the mesh size (or typical separation of neighboring filaments) $\xi$ and the cross--link separation $l_{\rm c}$. We observe from this strong dependence of the modulus on $l_{\rm c}$, which is independent of filament concentration in 3D, that the modulus of semi-flexible gels can be varied significantly (in fact, by orders of magnitude) with changes only in cross--link density at the same filament concentration. This is very different from the situation for flexible polymer gels, and may well be important for cells, in that the mechanical properties can be tuned by local variations in the densities or binding constants of various actin binding proteins. 

Furthermore, as we have shown \cite{head:03,wilhelm:03}, the modulus becomes a very strong function of concentration (which, again, will translate to cross--link density on 3D networks) in the nonaffine regime. In the nonaffine regime, the modulus can vary by several orders of magnitude with respect to the stiffer affine gels. As this nonaffine regime is expected for just a few (specifically, of order 10 or fewer) cross--links along a single filament, it may also be possible that the cell can reduce its stiffness significantly, and even fluidize, by decreasing the number of cross--links per filament or the filament length. In addition, by using its proximity to the $NA \longrightarrow AE$ cross-over, the cell can tune its nonlinear mechanical properties. In the $AE$ regime the cytoskeletal network should be strongly strain--stiffening due to the nonlinear extensional properties of individual filaments~\cite{mackintosh:95}. In the $NA$ regime, there should be a much larger linear regime since the bending modes of the filaments, which dominate the deformations in the $NA$ regime have a much larger linear response regime. Finally, we speculate that in the affine regime, the mechanical properties of the cell should be insensitive to the details of the cytoskeletal microstructure; in the $AE$ regime the mean-field character of the network enforced by the large ratio of $L$ to $\lambda$ suggests that local effects of cross-linker type or network topology to self-average. On the other hand, within the $NA$ regime, the cellular mechanical properties may be quite sensitive to such local network modifications. 

In this model, we have assumed freely-rotating crosslinks. In the case of actin networks, however, it is well know that many
associated proteins can bind actin filaments at either preferred or fixed angles~\cite{alberts:88}. This can have two distinct effects: one geometric, and the other mechanical. One the one hand, the model we have described is only for isotropic networks. Thus, if actin crosslinking results in an anisotropic network ({\em e.g.}, with oriented bundles), then one cannot describe such a system with the model presented here.

If, on the other hand, the networks remain isotropic, but with rigid bond angles between filaments, then we expect to see additional rigidity of the networks as a result. The size of this effect can be estimated for the affine regime (AE and AM, in fact). We find that the relative contribution to the network elastic modulus due to such crosslinks is only of order $l_{\rm b}/l_{\rm c}$, which is expected to be small for realistic networks of actin, both {\em in vitro} and {\em in vivo}. In simple terms, this is simply due to the very large lever arm that a filament segment of length $l_{\rm c}$ (say, of order 1 micrometer) between crosslinks has over a small a small actin binding protein of size a few nanometers. More precisely, this can be seen by noting that in shearing two filaments that cross at a finite angle in the shear plane, a fixed bond angle between the filaments will give rise to a distortion ({\em i.e.}, nonaffine) of the resulting filament conformations within a region $l_{\rm b}$ near the crosslink. (This length corresponds to the range over which finite bending occurs, {\em e.g.}, when a finite bending moment is imposed at a filament end.) The angle of this bend will be at most of order the macroscopic strain $\epsilon$ in the affine regime. This results in a bending elastic energy of order $\kappa\epsilon^2/l_{\rm b}$ per crosslink, compared with the longitudinal elastic energy of order $\mu\epsilon^2 l_{\rm c}$ per segment between crosslinks. Noting that $l_{\rm b}^{2}=\kappa/\mu$, we find that the latter term (which corresponds to the freely-rotating crosslink case) is larger than the former by a factor of order $l_{\rm b}/l_{\rm c}$.

For simple mechanical networks at $T=0$, as we consider in most of this paper, $l_{\rm b}$ is of order the molecular diameter, which is much smaller than the distance between overlapping filaments, let alone crosslinks in any real actin network. In the {\em in vitro} networks that have been studied, we expect this ratio to be no larger than at most a few percent. In the case of networks at finite temperature, as we discuss in Sec.~\ref{s:thermal}, $l_{\rm b}\sim\sqrt{l_{\rm c}^3/\ell_p}$, for which the ratio above is of order $\sqrt{l_{\rm c}/\ell_p}$. By definition, this is smaller than 1 for semiflexible networks. Thus, in any case, the corrections to the affine elastic moduli due to possible fixed-angle crosslinks (in isotropic networks) is expected to be small.

In the related studies by Wilhelm and Frey~\cite{wilhelm:03},
who also consider the $T=0$ mechanical properties of networks
such as ours,
the authors looked at both fixed-angle and freely-rotating crosslinks. They found that the rigidity percolation transitions occurred at somewhat different values of concentration for fixed-angle and freely-rotating bonds. But, they report that very similar behavior was observed for the two cases above the critical points. Specifically, they found no statistically significant difference in the dependence of the shear moduli with concentration in the non-affine elastic regime. Thus, it would appear that no substantial differences due to the mechanics of crosslinks can be expected in either affine nor nonaffine regimes, at least for the relatively sparse isotropic networks that actin forms.

\section*{Acknowledgements}

AJL and FCM would like to acknowledge D.A.\ Weitz for helpful conversations. AJL would like to acknowledge the hospitality of the Vrije Universiteit. DAH was partly funded by a European Community
Marie Curie Fellowship. This work is supported in part by the National Science
Foundation under Grant Nos. DMR98-70785 and PHY99-07949.

\appendix

\section*{Appendix}
\label{s:sims}

The filaments are deposited into the shear cell
as already described in Sec.~\ref{s:model}.
This is internally represented by the set of
points $\{{\bf x}_{i}\}$ consisting of all
cross--links and midpoints between cross--links
(the midpoints are included so that the
first bending mode between any two cross--links
is represented).
Relative motion between the ${\bf x}_{i}$ contributes
to the system Hamiltonian according to discrete
versions of (\ref{e:H_bend}) and (\ref{e:H_stretch}).
A change in separation
from $l_{0}$ to $l_{0}+\delta l$
between any two adjacently connected points
incurs an energy cost

\begin{equation}
\delta{\mathcal H}_{\rm stretch}
=
\frac{\mu}{2}
\left(
\frac{\delta l}{l_{0}}
\right)^{2}
l_{0}
\label{e:app1}
\end{equation}

\noindent{}In addition to this,
a non--zero angle $\delta\theta$
between the vectors ${\bf x}_{i}-{\bf x}_{i-1}$
and ${\bf x}_{i+1}-{\bf x}_{i}$\,,
where ${\bf x}_{i-1}$, ${\bf x}_{i}$ and ${\bf x}_{i+1}$
are consecutive adjacent points on the same filament,
contributes   

\begin{equation}
\delta{\mathcal H}_{\rm bend}   
=
\frac{\kappa}{2}
\left(
\frac{\delta\theta}{l'}
\right)^{2}
l'
\label{e:app2}
\end{equation}

\noindent{}where $l'$ is the mean of the
lengths to either side of the central point,
{\em i.e.}
$l'=\frac{1}{2}(|{\bf x}_{i}-{\bf x}_{i-1}|+|{\bf x}_{i+1}-{\bf x}_{i}|)$.
cross--linked filaments are coupled by imposing the
same ${\bf x}_{i}$ at intersections, but there is
no energy cost for relative angles between filaments:
cross--links can freely rotate.

Each contribution (\ref{e:app1}), (\ref{e:app2})
is linearised with respect to changes in
the ${\bf x}_{i}$ and summed to create the
system Hamiltonian ${\mathcal H}(\{{\bf x}_{i}\})$.
Either a uniaxial or shear strain $\gamma$ is applied
to the system through the periodic boundaries
in a Lees--Edwards manner~\cite{Lees_Edwards}.
The Hamiltonian ${\mathcal H}(\{{\bf x}_{i}\})$
is then minimized with respect to the $\{{\bf x}_{i}\}$
by the conjugate gradient method~\cite{conj_grad}.
Two optimizations are included.
The Hessian matrix
$A_{ij}=\partial^{2}{\cal H}/\partial x_{i}\partial x_{j}$
is preconditioned by $M^{-1}$, where $M$ has the same
diagonal $2\times2$ matrices of $A$ but is zero
elsewhere.
Furthermore, cross--links that lie within a given small distance,
typically $\approx10^{-3}L$, are coalesced.
This improves the conditioning of $A$ and hence the speed of
convergence considerably, while producing only minimal
change in the measured quantities, except precisely at
the transition.



\begin{thebibliography}{99}
\bibitem{ends}
Since dangling chain ends can carry no static stress, the volume fraction of such dangling ends serves only to dilute the cross-links. This effect can lead to corrections of at most $\frac{l_{\rm c}}{L}$. In three dimensions, it is also true that interchain steric interactions can play a role in determining the modulus of the material since entanglements can lead to topologically enforced cross-linking of the chains.
\end{thebibliography}
\end{document}